# Stimulated Raman phase shift spectroscopy: a pathway to hyperfine fingerprint spectra

Meng Qi,[1] Wenrong Wang,[1] Yuan-ji Li [1,2*], Jin-xia Feng [1,2], and Kuan-shou Zhang[1,2*]

**Abstract**

The principle and experimental realization of a novel Raman spectroscopic technique entitled stimulated Raman phase shift (SRPS) spectroscopy was demonstrated. This technique depends on the measurement of the stimulated Raman scattering (SRS) induced phase shift of Stokes light field ($\Delta\varphi$) that is related to the real part of the third order nonlinear susceptibility of SRS. In principle, the spectral lineshape of $1/|\Delta\varphi|$ is a delta function waveform, which is insensitive to the fluctuation of Stokes light intensity, the decoherence of phonon in materials, as well as the inhomogeneous fluorescence background. In order to measure $1/|\Delta\varphi|$, a SRPS spectrometer including a Mach-Zender interferometer and a signal processing device was developed. Using the developed spectrometer, the SRPS and stimulated Raman gain (SRG) spectra of neat dimethyl sulfoxide were detected simultaneously. Seven Raman peaks corresponding to specific molecule vibrational and rotational modes were observed in the SRPS spectra, while only two peaks could be identified in the SRG spectra without a priori knowledge. The linewidth of the Raman peak centered at 2913.283 cm$^{-1}$ indicating the $v_s$(CH$_3$) stretching mode of the methyl groups was less than 0.00036 cm$^{-1}$ in the measured SRPS spectra, which was almost four orders of magnitude narrower than that in the measured SRG spectra. Meanwhile, the detection signal-to-noise ratio of the Raman peak centered at 2913.283 cm$^{-1}$ was 25.3 dB, representing an increase of 14.3 dB compared to the SRG spectra. The reliability of SRPS technique was verified by 10 independent measurements, and the standard deviation of the Raman peak frequency was less than ±0.338 cm$^{-1}$. The SRPS technique paves the way for characterizing the hyperfine fingerprint of materials.

## Introduction

Raman scattering (RS) spectroscopy has emerged as a powerful technique for obtaining molecular vibrational and rotational information with high sensitivity and specificity [1-5]. In comparison with other methods, e.g. infrared absorption spectroscopy and fluorescence spectroscopy, RS provides sharper spectral peaks, lower detection limit, and quantitative analysis capability. As a consequence, RS becomes the preferred approach in many important applications, such as rapid detection of COVID-19 virus without polymerase chain reaction [6, 7], early detection of bio-markers for cancer and nervous system diseases [8-10], etc.

In previous researches on RS, much more attention had been focused on optimizing detection signal-to-noise ratio (SNR) via raising up the RS signal or reducing the background noise. Generally, RS signal enhancement relied on either local-plasmon-field enhancement, for instance surface-enhanced RS (SERS) [11-14] and tip-enhanced RS (TERS) [15, 16]; or coherent enhancement, namely stimulated Raman scattering (SRS) [17, 18] and coherent anti-stokes Raman spectroscopy (CARS) [19, 20]. Recently, the tracking of single chemical bond and its movement such as tilting and later hopping in a CO molecular had been demonstrated using TERS [16]. To reduce the background noise, orthogonal polarization SRS [21, 22] and quantum enhanced SRS [23, 24] had been developed. By making two Pump and Stokes laser pulse pairs in perpendicular polarization, where each of them acts as an intensity reference for the other, SRS spectra detection with nearly shot-noise limit SNR can be achieved using a fiber-based femtosecond laser system [22]. Furthermore, when a bright squeezed state was employed as Stokes light replacing the coherent light field, the background noise originated from the laser intensity noise can be reduced, leading to a 3.6 dB enhancement of SNR in comparison with continuous wave (CW) SRS spectroscopy [23].

However, in some application scenarios, precise detection of Raman shift is a more pivotal factor. As discussed in [25-26], characteristic spectral frequencies of a chemical bond may experience frequency deviations once the ambient condition are changed. If the peak linewidth of RS spectra is narrow enough and the frequency detection uncertainty of the spectrometer is excellent, accurate calibration of Raman spectrometer, hyperfine spectroscopy, and multi-parameter synchronous sensing can be achieved. Moreover, discrimination of different kinds of molecules with minor structure differences is also urgently needed in fine chemistry, early diagnosis of diseases, etc. Unfortunately, the raw data of the Raman spectra obtained using the existing methods suffered from the spectral broadening induced by laser linewidth, decoherence of phonon in materials, as well as the inhomogeneous fluorescence background. Consequently, processing including baseline correction, peak search and Lorentz curve fitting were employed to extract the Raman peak information [17-22]. However, during these processing, large errors of extracted Raman peak frequencies may be introduced, especially in the case that multiple Raman peaks are very close to each other [21, 22].

[1]State Key Laboratory of Quantum Optics and Quantum Optics Devices, Institute of Opto-Electronics, Shanxi University, Taiyuan 030006, China
[2]Collaborative Innovation Center of extreme Optics, Shanxi University, Taiyuan 030006, China
Correspondence: Kuanshou Zhang (kuanshou@sxu.edu.cn) and Yuanji Li (liyuanji@sxu.edu.cn)

Up to date, the spectral broadening induced by laser linewidth had been well reduced using a SRS spectrometer based on CW single frequency lasers, which exhibited the best measured frequency resolution of 0.07 cm$^{-1}$ and detection step of 0.9 cm$^{-1}$ [27], two or more magnitudes better than that employing femtosecond or picosecond laser sources [28, 29]. Nevertheless, since further compressing of laser linewidth has little significance, the Lorentz type spectral broadening of Raman peaks and fluorescence background become the dominators restricting the Raman peak linewidth and frequency resolution of RS.

Similar to other third order nonlinear optical process, CARS and SRS not only varied the optical field intensity that was related to the imaginary part of third order nonlinear susceptibility ($\chi''_R$), but also introduced nonlinear phase shift related to the real part of the nonlinear susceptibility ($\chi'_R$). Several studies had been demonstrated the experimental measurements of phase shift induced by CARS [30-37] or SRS [37-38], in which the SNR of CARS spectra can be improved since the phase shift signal was not disturbed by the nonresonant background, and the phase shift information were usually used to modify the spectral shape of CARS or construct multi-dimensional Raman spectra composed with the measured imaginary part and real part of the third order nonlinear susceptibility. However, the Lorentz type spectral broadening of Raman peaks was still existed in the present results in these studies, and the information of Raman induced phase shift was not sufficiently developed, i.e. the phase shift information was only used as a supplement or a replacement with no more information for the SRG spectra.

Here, we propose a novel stimulated Raman phase shift (SRPS) spectroscopic technique for characterizing the hyperfine fingerprint of materials. The mechanism of the technique is based on detecting and exploring the complete information of the SRS induced phase shift rather than the SRS induced light intensity variation. Using the developed SRPS spectrometer, quantitative RS spectra without Lorentz type Raman spectra broadening and fluorescence background can be obtained.

## Principle of SRPS spectroscopy

Generally, the SRS spectra were obtained by detecting the stimulated Raman gain (SRG) [21, 39], i.e. the intensity difference of Stokes light induced by SRS, or stimulated Raman loss (SRL) [17,40,41], i.e. the intensity difference of Pump light induced by SRS, and tuning the Pump wavelength. The signal intensity of SRG near the Raman resonance can be written as [42]

$$I_{SRG} = I_{s0}\left(\exp G_0 - 1\right),$$
$$G_0 = \frac{N I_p \alpha^2 L}{c^2 n_p n_s m_0 \omega_v} \frac{\omega_s \gamma}{[(\omega_v - \Delta\omega)^2 + \gamma^2]}, \Delta\omega = \omega_p - \omega_s, \quad (1)$$

where $I_{s0}$ and $\omega_s$ are respectively the initial intensity and angular frequency of Stokes light, $I_p$ and $\omega_p$ are respectively the initial intensity and angular frequency of Pump light, $n_s$ and $n_p$ are the refractive indices of the sample at the wavelengths of Stokes and Pump, $L$ is the thickness of the sample, $c$ is the light speed in vacuum, $\varepsilon_0$ is the permittivity of vacuum, $\omega_v$ is the resonant angular frequency of the molecular vibrational and rotational mode, $\alpha$ is the partial differential of the nonlinear polarizability of the dipole corresponding to the molecular mode, $\gamma$ is the normal mode dephasing time, $N$ is the number of molecules inside the scattering volume, $m_0$ is the molecular reduced mass.

According to the definition of refractive index, the variation of refractive index ($\Delta n$) induced by SRS can be read as

$$\Delta n \approx \frac{\chi'_R I_p}{\varepsilon_0 c n_p n_s}. \quad (2)$$

Then the signal intensity of the SRS induced phase shift ($\Delta\varphi$) and its reciprocal of absolute value ($1/|\Delta\varphi|$) can be read as

$$\Delta\varphi \approx \frac{\omega_v - \Delta\omega}{(\omega_v - \Delta\omega)^2 + \gamma^2} \cdot \frac{N \alpha^2 I_p L}{12 m_0 \omega_v c n_p n_s}, \quad (3)$$

$$\frac{1}{|\Delta\varphi|} \approx \left|\frac{(\omega_v - \Delta\omega)^2 + \gamma^2}{\omega_v - \Delta\omega}\right| \cdot \frac{12 m_0 \omega_v c n_p n_s}{N \alpha^2 I_p L}, \quad (4)$$

And the signal intensity of SRPS can be defined as

$$I_{SRPS} = \left[\frac{1}{|\Delta\varphi|}\right]_{Normalized} \cdot \left(\frac{\partial \Delta\varphi}{\partial \omega_p}\right)_{max}$$
$$\approx \left[\frac{1}{|\Delta\varphi|}\right]_{Normalized} \cdot \frac{N \alpha^2 I_p L}{12 m_0 \omega_v c n_p n_s} \frac{1}{\gamma^2}. \quad (5)$$

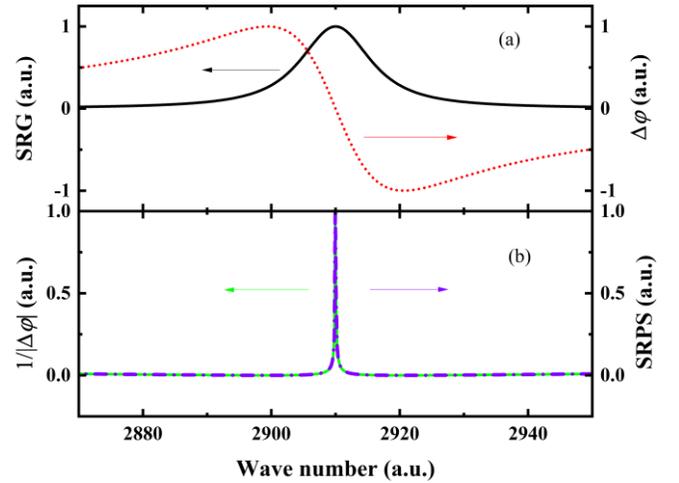

**Fig. 1.** Normalized theoretical simulation results of (a) SRG spectra and $\Delta\varphi$-versus-$\Delta\omega$ spectra, (b) (1/|$\Delta\varphi$|)-versus-$\Delta\omega$ spectra and SRPS spectra.

To make a comparison between SRG, $\Delta\varphi$-versus-$\Delta\omega$, (1/|$\Delta\varphi$|)-versus-$\Delta\omega$ and SRPS spectra, theoretical simulations were carried out using Eqs.(1),(3)-(5), and setting the parameter values as $\omega_s$=1.77*10$^{15}$ Hz, $\omega_v$=5.48*10$^{14}$ Hz, $n_s$=1.5, $n_p$=1.4, $I_p$=3.2*10$^6$ W/m$^2$, $\gamma$=2*10$^{12}$ Hz, $\alpha^2$ =4.65*10$^{-23}$ C·m/V, $N$ =5*10$^{22}$, $m_0$=3*10$^{-26}$ kg, $L$=1 mm. The normalized simulation results of the four spectra around the Raman peak frequency of $\omega_v$ are depicted in Fig.1(a) and 1(b). For SRG spectra (solid curve in Fig.1(a)), Lorentz type spectral broadening stemmed from decoherence of phonon is observed around $\omega_v$, while the linewidth is about 13.08 cm$^{-1}$. For $\Delta\varphi$-versus-$\Delta\omega$ spectra (dot curve in Fig.1(a)), which were introduced in Ref[30-38], shows an antisymmetry lineshape with Lorentz type

spectral broadening that can be employed to suppress linear and nonlinear backgrounds with the help of balanced-detection method. For the normalized ($1/|\Delta\varphi|$)-versus-$\Delta\omega$ and SRPS spectra (solid and dash-dot curves in Fig.1(b)), the lineshapes are identical, and the linewidth of the curve is significantly compressed leading to a delta-function waveform with the peak located at $\omega_v$. This phenomena can be understood when one notice that the absolute value of the real part of $\chi_R$ is 0 at the eigen-frequencies of the molecular vibrational and rotational modes, i.e. the Raman peak frequency, but that is a finite real number at other neighboring frequencies. Therefore, the value of $1/|\Delta\varphi|$ at the Raman peak frequency is far beyond that at other frequencies, leading to a SNR approaching infinity and a linewidth approaching 0. Apparently, if the SRS induced optical field phase shift can be precisely detected, the novel SRPS spectra shown in Fig.1(b) will possess multiple advantages in comparison with the SRG spectra and common phase spectra [30-38]. Firstly, the linewidth of the spectral peak becomes quite narrow, so that the fine variation of Raman peak frequency induced by ambient condition change can be monitored precisely. Secondly, since the phase change is not relevant to the Stokes intensity, and the SNR is approaching infinity in principle, non-invasive high sensitivity RS spectra detection is allowed without the help of Raman probe or SERS substrate. Thirdly, the SRPS spectra is proportional to $NL$ corresponding to the molecular number and $\alpha^2$ corresponding to the Raman response behavior of the specific vibrational and rotational mode. Thus the complete quantitative fingerprint information can be obtained from only phase spectra.

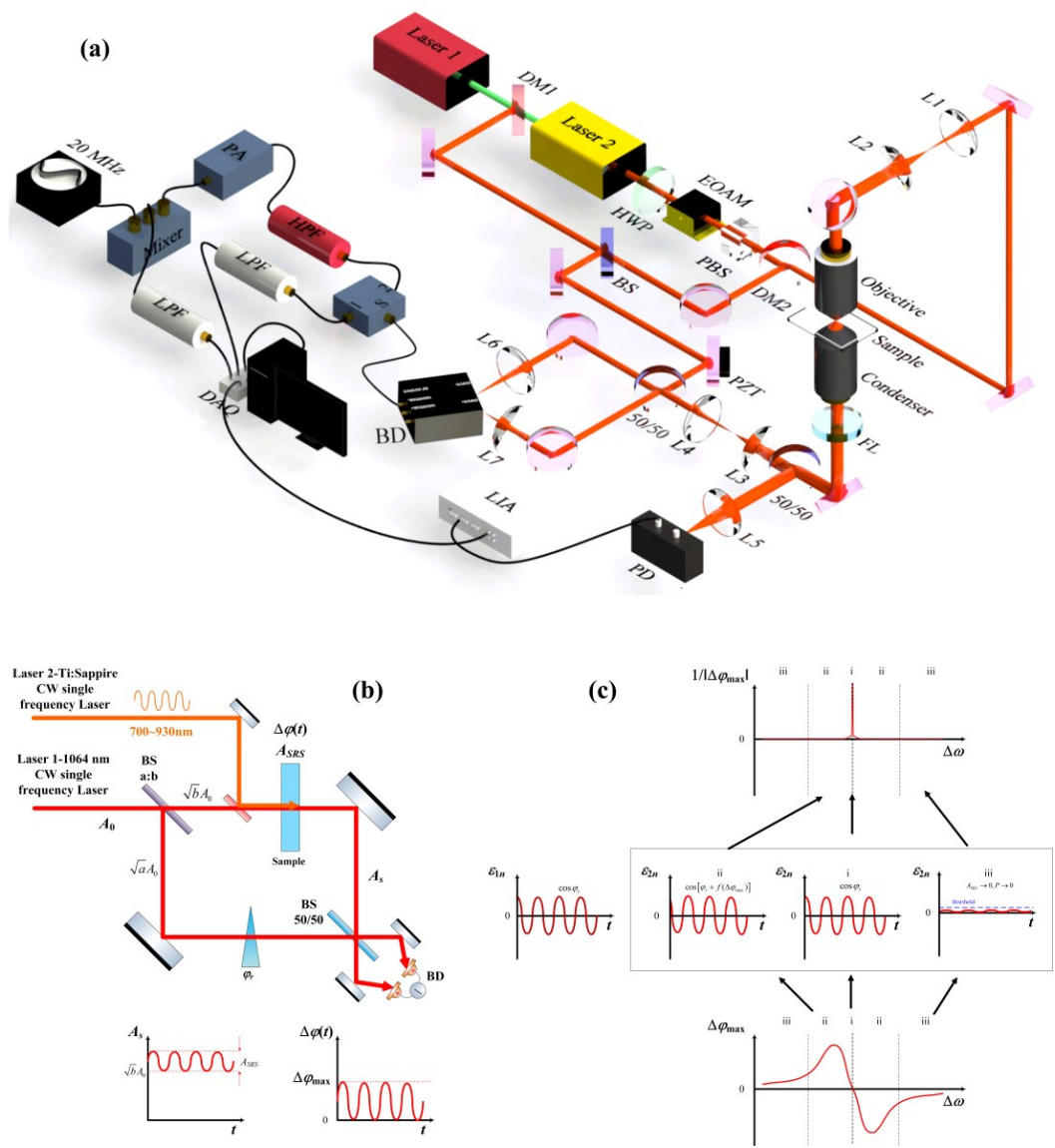

**Fig. 2.** (a) Schematic of the SRPS spectrometer. Laser 1: CW single frequency 1064 nm/532 nm dual-wavelength laser; Laser 2: Ti-sapphire CW single frequency tunable laser; DM1-2: dichroic mirror; HWP1-2: half-wave plate; EOAM: electro-optic amplitude modulator; PBS: polarized beam splitter; BS: beam splitter; FL: long-wavelength pass optical filter; 50/50: 50/50 beam splitter; L1-7: Lens; PD: photodetector; LIA: lock-in amplifier; BD: balanced detector; RS: radio frequency splitter; HPF: high pass filter; LPF: low pass filter; PA: power amplifier; SG: signal generator; DAQ: data acquisition device. (b) Schematic for the interferometer. (c) Principle for signal processing.

To measure the phase shift induced by SRS, a SRPS spectrometer setup was designed and built, as shown in Fig. 2. The 1064 nm laser from a home-made CW single frequency 1064 nm and 532 nm dual-wavelength laser was employed as Stokes light [43], which was split into the local Stokes light and the injected Stokes light by a beam splitter (BS) with a split ratio of a:b. A home-made Ti-sapphire CW single frequency tunable laser excited by the 532 nm laser from the dual-wavelength laser was used as Pump light [44]. The Pump light that was intensity modulated using an electro-optic amplitude modulator (EOAM, Thorlabs, Model: EO-AM-R-20-C1) driven by a signal generator (SG, RIGOL, Model: DG5102) and the injected Stokes light were confocused on the sample via an objective (Olympus, Model: NIR Apo) and the Stokes light from the sample was collected by a condenser (Olympus, Model: U-AAC). The amplitude of the collected Stokes light and the phase shift induced by SRS can be written as

$$A_s = \sqrt{b}A_0 \exp[-i\Delta\varphi(t)] + A_{SRS}\left[\frac{1+m\cos(\omega_m t+\varphi_0)}{2}\right]\exp[-i\Delta\varphi(t)], \quad (6)$$

$$\Delta\varphi(t) = \frac{\Delta\varphi_{max}}{4}\left[\frac{m^2+2}{2} + 2m\cos(\omega_m t) + \frac{m^2}{2}\cos(2\omega_m t)\right], \quad (7)$$

$$\Delta\varphi_{max} = \frac{\omega_v - \omega_p + \omega_s}{(\omega_s - \omega_p + \omega_v)^2 + \gamma^2} \cdot \frac{N\alpha^2 I_{p0} L}{12m_0\omega_v c n_p n_s}. \quad (8)$$

where $A_0$ is the total amplitude of initial Stokes light, $A_{SRS}$ is the maximum amplitude difference of Stokes light due to the amplification via SRS. $I_{p0}$ is the maximum pump intensity during modulation, $m$ is the modulation depth, $\omega_m$ is the modulation frequency, $\varphi_0$ is the initial phase of modulation signal, $\Delta\varphi_{max}$ is the maximum phase shifty induced by SRS during intensity modulation. The collected Stokes light was split by a 50/50 beam splitter. One part of the collected Stokes light was detected using a home-made photodetector (PD). The SRG spectra was directly derived by demodulating the output of PD using a lock-in amplifier (LIA, Zurich Instrument, Model: HF2LI).

To measure the SRPS spectra simultaneously, we constructed a Mach-Zender interferometer, in which the Stokes light was split into the local Stokes light and the injected Stokes light by the BS, the local Stokes light and the residual part of the collected Stokes light were interfered on the 50/50 beam splitter. The interference lights were detected by a balanced detector (BD, Thorlabs, Model: PDB450C) composed with two identical photodiodes and a self-subtraction trans-impedance amplifier circuits. Supposing that $m\approx1$, the difference signal from the BD can be read as

$$I_- = \sqrt{a}A_0 \left\{ \begin{array}{l} 2\sqrt{b}A_0\left[\cos\varphi_r - \sin\varphi_r\Delta\varphi(t)\right] \\ + A_{SRS}\left[1+\cos(\omega_m t+\varphi_0)\right]\left[\cos\varphi_r - \sin\varphi_r\Delta\varphi(t)\right] \end{array} \right\}, \quad (9)$$

where $\varphi_r$ is the relative phase delay between the local Stokes light and the collected Stokes light.

To extract the phase shift signal from Eq.(9), the difference signal from BD was split to be two parts with identical radio frequency (RF) power using a RF splitter (RS, Mini-circuits, Model: ZFRSC-42-S). The two signals were processed to get the low-frequency term and high-frequency term of the difference signal, respectively. Considering that $A_0 \gg A_{SRS}$ and $\chi'_R \ll 1$, the low-frequency term obtained using a low-pass filter (LPF, Mini-circuits, Model: BLP-1.9+) can be expressed as

$$\varepsilon_1 = \sqrt{ab}A_0^2 \cos\varphi_r. \quad (10)$$

The high-frequency term was derived and amplified via a high pass filter (HPF, Thorlabs, Model: EF515) and a power amplifier (PA, Mini-circuits, Model: ZHL-1-2W-S+). The signal was then mixed with a demodulation signal using a mixer (Mini-circuits, Model: ZAD-1-1+). When the demodulation signal is $B\cos(\omega_m t+\varphi_0)$, the mixing signal after LPF can be expressed as

$$\varepsilon_2 = B\sqrt{a}A_0\left[A_{SRS}\cos\varphi_r/2 - \sqrt{b}A_0\Delta\varphi_{max}\sin\varphi_r\cos\varphi_0\right], \quad (11)$$

Obviously, when the frequency difference between Stokes and Pump ($\omega_p-\omega_s$) is equal to $\omega_v$, the second term in Eq.(11) is zero, so that the two signals $\varepsilon_1$ and $\varepsilon_2$ have the same phase. When $\omega_p-\omega_s$ is far away from $\omega_v$, both $A_{SRS}$ and $\Delta\varphi_{max}$ approach zero, leading to a nearly constant phase difference between $\varepsilon_1$ and $\varepsilon_2$. When $\omega_p-\omega_s$ is slightly deviated from $\omega_v$, the $\varepsilon_2$ signal comes to be

$$\varepsilon_2 = \sqrt{g^2+h^2}\cos(\varphi_r+q),$$
$$g = B\sqrt{a}A_0 A_{SRS}/2, \quad h = \sqrt{ab}BA_0^2\Delta\varphi_{max}\cos\varphi_0, \quad (12)$$
$$q = \arctan(h/g).$$

The phase shift can be obtained using Eq. (13)

$$\Delta\varphi_{max} = \tan\left[\arccos\left[\varepsilon_{1n}\varepsilon_{2n} + \sqrt{(1-\varepsilon_{1n}^2)(1-\varepsilon_{2n}^2)}\right]\right]. \quad (13)$$

where $\varepsilon_{1n}$ and $\varepsilon_{2n}$ are the normalized value of $\varepsilon_1$ and $\varepsilon_2$. And the partial differential $\partial\Delta\varphi_{max}/\partial\omega_p$ can be derived using the phase shift at the tested $\omega_p$ and that at the adjacent two $\omega_p$.

The $1/|\Delta\varphi_{max}|$ signal can be obtained using Eq. (14)

$$\frac{1}{|\Delta\varphi_{max}|} \propto \left\{ \begin{array}{ll} 1/\tan\left[\left|\arccos\left[\varepsilon_{1n}\varepsilon_{2n} + \sqrt{(1-\varepsilon_{1n}^2)(1-\varepsilon_{2n}^2)}\right]\right|\right], & \varepsilon_2 \neq 0 \\ 0, & \varepsilon_2 = 0 \end{array} \right. . \quad (14)$$

Note that the case that $\omega_p-\omega_s$ is far away from $\omega_v$ is identified according the value of $\varepsilon_2$. Finally the SRPS signal can be obtained by multiply $1/|\Delta\varphi|$ with $\partial\Delta\varphi/\partial\omega_p$.

**Results**

around 2995.162 cm$^{-1}$ can be observed without a priori knowledge. Usually, these two modes can be respectively assigned to the $v_s$(CH$_3$) and $v_a$(CH$_3$) modes of the methyl groups [31]. One can also found that the full width at half maximum (FWHM) of the Raman peak at 2914.253 cm$^{-1}$ is 14.35 cm$^{-1}$, while that for the other one is 10.01 cm$^{-1}$. In the SRPS spectra (processing procedure seen in Method and supplement), the Raman spectra came to be a combination of a series of very sharp peaks, where the signal strengths indicate the value of $\partial\Delta\varphi/\partial\omega_p$ at the resonance frequency. Beside the two peaks centred at 2913.283 cm$^{-1}$ and 2997.594 cm$^{-1}$ that correspond to the $v_s$(CH$_3$) and $v_a$(CH$_3$) modes of the methyl groups, one can also found a peak located at 2875.778 cm$^{-1}$ that can be assigned to the overtones of the methyl rocking and deformation modes [40]. The measured SRPS spectra exhibited a maximum SNR of 25.3 dB, which was 14.3 dB higher than that of the measured SRG spectra. Furthermore, it is worth noting that the FWHM of the Raman peak centred at 2913.283 cm$^{-1}$ in the measured SRPS spectra was 0.00036 cm$^{-1}$ (shown in the inset in Fig.3(b)), almost four magnitudes narrower than that in the measured SRG spectra. This value was restricted by the stability of the detection system and the tuning interval of the Pump wavelength.

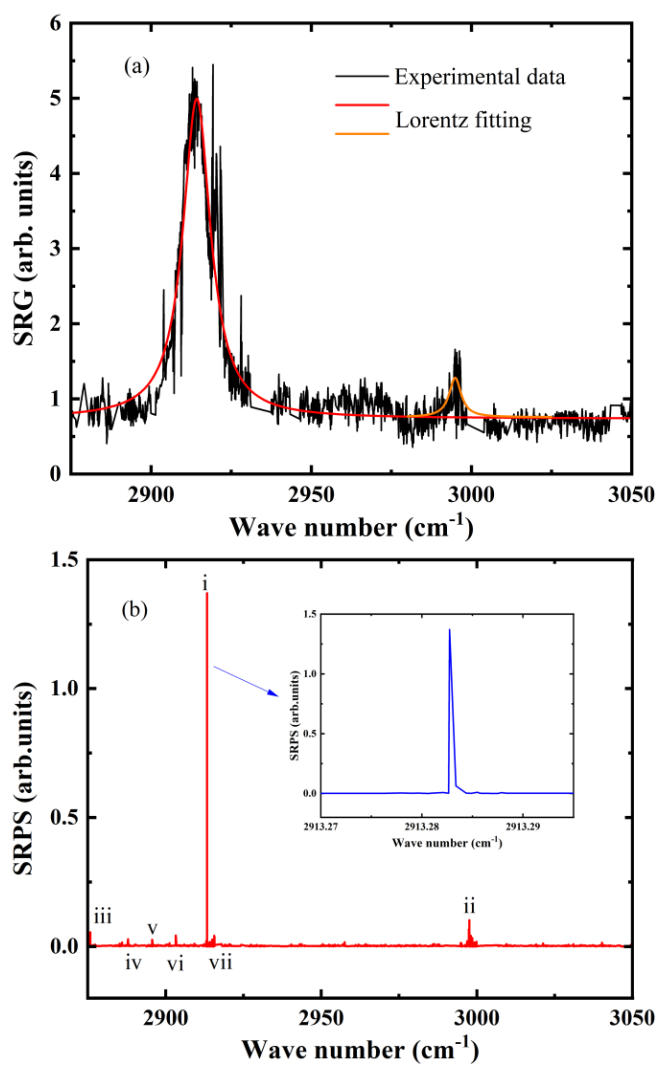

**Fig. 3.** (a) SRG spectra of DMSO; Lorentz fitting: Peak fitting using Lorentz function; (b) SRPS spectra of DMSO. i: Raman peak corresponding to $v_s$(CH$_3$) mode of the methyl groups; ii: Raman peak corresponding to $v_a$(CH$_3$) mode of the methyl groups; iii: Raman peak corresponding to the overtones of the methyl rocking and deformation modes. Inset: Detail view of the SRPS spectra of DMSO around Raman peak i.

Fig.3(a) and 3(b) shows the measured SRG and SRPS spectra of neat dimethyl sulfoxide (DMSO) when the powers of the Stokes light incident onto the sample and the local Stokes light were 20 mW and 7 mW, and the Pump light power before EOAM was maintained at 80 mW by a half-wave plate (HWP2) fixed on an electric rotary holder and a polarized beam splitter (PBS) during the Pump wavelength tuning from 700 nm to 930 nm with a tuning step of about 0.001 nm (the step was not a constant due to the nonlinear response of the tuning device composed with the birefringent filter and etalon that were fixed on two rotating stepper motor, the minimum measured step reach 6.59*10$^{-6}$ nm, wavelength monitoring was realized using a commercial wavelength meter and a Fabry-Perot interferometer with a finesse of 280. In the SRG spectra, peak search and Lorentz curve fitting are employed to extract the Raman peak information, one strong peak (i) centred at the wave number of 2914.253 cm$^{-1}$ and a weak peak (ii) located

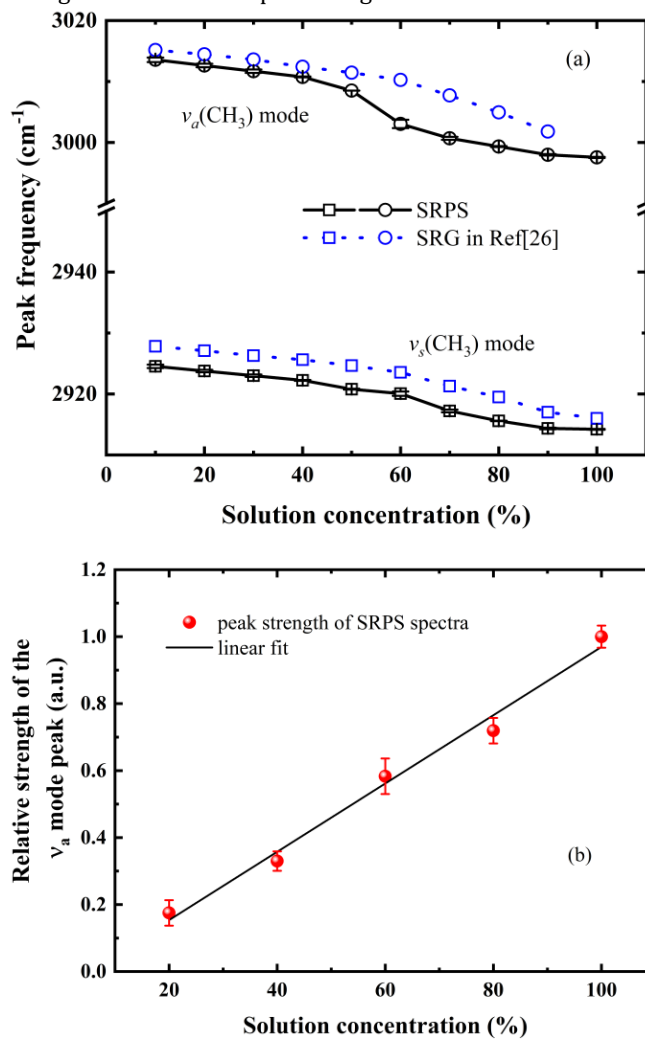

**Fig. 4.** (a) Peak frequency of the measured SRG spectra (i), SRPS spectra (ii), and that of the SRG spectra reported in Ref[26] (iii); (b) signal strength of the peak signal in SRPS spectra as a function of DMSO-water solution concentration.

As mentioned in Ref [26], the Raman peak frequency always experiences a deviation when DMSO is dissolved in water, and the deviation varies with the change of the concentration of DMSO. Figure 4(a) shows the behavior of characteristic frequencies of the $v_s(CH_3)$ and $v_a(CH_3)$ modes measured by the SRPS spectrometer versus solution concentration. The averaged peak frequency and the corresponding standard deviation were obtained by 10 repeated measurements. The evolution tendencies of the two curves agree well with the results in Ref [26], while the averaged peak frequency and the corresponding standard deviation of the $v_s(CH_3)$ mode at the concentrations ranging from 10% to 100% are 2924.5350 cm$^{-1}$(±0.139 cm$^{-1}$), 2923.7654 cm$^{-1}$(±0.101 cm$^{-1}$), 2922.9957 cm$^{-1}$(±0.101 cm$^{-1}$), 2922.2261 cm$^{-1}$(±0.031 cm$^{-1}$), 2920.7925 cm$^{-1}$(±0.037 cm$^{-1}$), 2920.0544 cm$^{-1}$(±0.176 cm$^{-1}$), 2917.2212 cm$^{-1}$(±0.112 cm$^{-1}$), 2915.5890 cm$^{-1}$(±0.019 cm$^{-1}$), 2914.3446 cm$^{-1}$(±0.064 cm$^{-1}$), 2913.283 cm$^{-1}$(±0.032 cm$^{-1}$); and that of the $v_a(CH_3)$ mode are 3013.5645 cm$^{-1}$(±0.185 cm$^{-1}$), 3012.6287 cm$^{-1}$(±0.145 cm$^{-1}$), 3011.6928 cm$^{-1}$(±0.138 cm$^{-1}$), 3010.7569 cm$^{-1}$(±0.032 cm$^{-1}$), 3008.5116 cm$^{-1}$(±0.001 cm$^{-1}$), 3003.0463 cm$^{-1}$(±0.338 cm$^{-1}$), 3000.6811 cm$^{-1}$(±0.122 cm$^{-1}$), 2999.3354 cm$^{-1}$(±0.040 cm$^{-1}$), 2997.9944 cm$^{-1}$(±0.061 cm$^{-1}$), 2997.594 cm$^{-1}$(±0.045 cm$^{-1}$). The standard deviations are less than ±0.338 cm$^{-1}$, which is nearly 1/3 of the results derived from our measured SRG spectra. Moreover, Figure 4(b) shows the signal strength of the peak signal in SRPS spectra ($\partial\Delta\varphi/\partial\omega_p$) as a function of DMSO-water solution concentration. A linear function y=-0.05x+0.0102 can be used to characterize the evolution behavior, which is nearly identical with the results obtained from the measured SRG spectra.

**Discussion**

In conclusion, we demonstrated the principle and experimental realization of a novel Raman spectroscopic technique entitled SRPS. The SRS induced phase shift was measured by developing a SRPS spectrometer incorporating a Mach-Zender interferometer and a signal processing device. The measured SRPS spectra exhibited multiple advantages, such as narrow linewidth of fingerprint peaks, high detection SNR, and good frequency detection uncertainty. For a neat DMSO sample, the measured SRPS spectra showed a FWHM of the fingerprint peak centered at 2913.283 cm$^{-1}$ as narrow as 0.00036 cm$^{-1}$, which was four magnitudes narrower than that in the measured SRG spectra. Owing to the background free behavior, the maximum SNR of the SRPS spectra reached 25.3 dB, which was 14.3 dB higher than that of the SRG spectra. For a series of DMSO-water solution, the evolution of the Raman peak frequencies along with varied solution concentration was detected using the SRPS spectrometer. The reliability of the SRPS method was verified by 10 independent measurements, and the standard deviation of the Raman peak frequency was less than ±0.338 cm$^{-1}$. The SRPS technique paves the way for characterizing the hyperfine fingerprint of materials, thus can be applied in the fields such as in-situ functional imaging of living organisms, hyperfine chemistry investigation, early diagnosis of diseases based on precise biomarkers detection, etc.